\documentclass[journal]{aa}
\usepackage{epsfig}
\usepackage{amsmath}
\usepackage{graphicx}
\usepackage{amssymb} 
\usepackage{pdflscape}
\usepackage{txfonts}
\usepackage{graphicx}
\usepackage{natbib}

\begin{document}

\title{Abundance anomalies in low mass field mergers as evidence of a merger origin for the second generation stars in globular clusters}

      \author{V. Kravtsov\inst{1}
          \and
              S. Dib\inst{2}
          \and
              F. Calderon\inst{3}              
}

\offprints{V. Kravtsov}

\institute{Sternberg Astronomical Institute, Lomonosov Moscow State University, Universitetsky pr. 13,
              119234 Moscow, Russia\\
             \email{vkravtsov1958@gmail.com}
      \and
              Max Planck Institute for Astronomy, K\"{o}nigstuhl 17, D-69117, Heidelberg, Germany\\
              \email{sami.dib@gmail.com}
      \and
              Departamento de F\'isica, Universidad Cat\'olica del Norte,
              Avenida Angamos 0610, Antofagasta, Chile\\
              \email{fcalderon@ucn.cl}                      
            }

   \date{Received xxxxx / Accepted xxxxx}

   \abstract
{The canonical formation of second-generation (G2) stars in globular clusters (GCs) from gas enriched and ejected by G1 (primordial) polluters faces substantial challenges, i.e. (i) a mass-budget problem and (ii) uncertainty in the source(s) of the abundance anomaly of light elements (AALE) in G2 stars. The merger of G1 low-mass main-sequence (MS) binaries can overcome (i), but its ability to result in AALE is omitted.} {We provide evidence of the merger process to explain AALE by relying on highly probable merger remnants in the Galactic disk.} {We focus on carbon-deficient red-clump giants with low mass of $1.0 M_{\sun} < M \lesssim 2.0 M_{\sun}$ and hot He-intermediate subdwarfs of supersolar metallicity, both manifesting G2-like AALE incompatible with GC origin.} {The origin of such rare core He-burning stars as the mergers of [MS star (MSS)]$+$[helium white dwarf (HeWD)] binaries, evolved from low-mass high-mass ratio (MSS$+$MSS) ones, is supported by models evolving to either horizontal branch (HB) stars or He subdwarfs via the red giant branch (RGB). Such binaries in the GC NGC 362 contain very young ($\sim$ 4 Myr) extremely low-mass HeWDs, in contrast to much older ($\sim 100$ times) counterparts in open clusters. This agrees with the impact of GC environment on the lifetime of hard binaries: (MSS$+$HeWD) systems merge there soon after arising from (MSS$+$MSS) binaries that underwent the common-envelope stage of evolution. From the number and lifetime of the (MSS$+$HeWD) binaries uncovered in NGC 362, the expected fraction of their progeny RGB G2 stars is estimated to be $\lesssim$10\%.} {The field merger remnants with G2-like AALE support the merger nature of at least a fraction of G2 stars in GCs. The specific channel [(MSS$+$MSS) -- (MSS$+$HeWD) -- merger product] supported by observations and models is tentatively identified as the channel of formation of the extreme G2 RGB component in GCs.}  {}


\titlerunning{The merger origin of G2 stars in globular clusters}
\authorrunning{Kravtsov et al.}

\maketitle

\section{Introduction}
\label{introduc}

Globular clusters (GCs) are known to be composed of stellar populations that are not quite simple. Two sub-populations or generations of stars, a primordial (G1) and secondary (G2) ones, are typically distinguished in GCs \citep{Carretta2009,Milone_etal2017} with a present-day mass exceeding a threshold level of about $M_{\rm GC} \sim 10^{4}$ M$_{\sun}$ \citep[e.g.,][]{Carrettaetal2010a,Bragagliaetal2017,Simpsonetal2017,Tangetal2021}. The surface abundance patterns of G2 stars, mainly belonging to the red giant branch (RGB)\footnote{These stars were studied first, most reliably, in large number of GCs.}, 
exhibit anomalies in the abundance of light, proton-capture elements in such a way that N, Na, (Al) are in excess and C, O, (Mg) in deficit. In contrast, G1 red giants and their field counterparts of the same metallicity are indistinguishable in this context. This is referred to as multiplicity in GCs. The most common spectroscopic indicator of the multiplicity in GCs is the Na-O anti-correlation among RGB stars \citep[e.g.,][and references therein]{Carretta2009} \footnote{Variations in the abundance of several light chemical elements and their molecules (e.g., CN, CH, NH) among RGB stars within individual GCs have been already known since the seventies of the last century \citep[see, for example, a review by][and references therein]{Kraft1979}.}. A particularly point of interest is that the Na-O anti-correlation was primarily spectroscopically observed in GCs among brighter stars, from the upper RGB down to the main-sequence (MS) turnoff point. So, it is not entirely clear whether the same Na-O anti-correlation is manifested by MS stars of lower luminosity and down to which limit. The Mg-Al anti-correlation is more controversial, because it tends to disappear in most metal-rich and\slash or least massive GCs \citep{Pancinoetal2017,Nataf_etal2019}. Most of the detections and quantitative studies of multiple populations in GCs or in intermediate-age massive star clusters, particularly in distant ones, are based on photometric effects that primarily arise from the characteristic abundances of CNO elements.

Inferring on multiple stellar populations in GCs broadly implied the canonical formation of G2 stars in the usual way, from the gas primarily ejected from G1 polluters (irrespective of their kind) and possibly diluted with pristine gas. However, it remains a contradictory issue being considered in the framework of the canonical approach. Indeed, several lines of evidence have converged, indicating that this approach faces serious challenges. First, perhaps the most problematic is the so-called mass-budget problem: a deficit of the ejected enriched gas necessary to form the observed fraction of G2 stars, especially in massive GCs, where the fraction of G2 stars on the RGB was found to be very high, up to 80\% \footnote{As a possible exception, see the scenario proposed by \citet{Bastianetal2013} who invoked the so-called "tail-end" accretion of the low-velocity gas expelled from massive binaries by the disks of low-mass pre-main-sequence stars.}. Second, the origin of the above-mentioned abundance anomaly in G2 stars is uncertain up to now. In principle, all conceivable polluters, the sources of the anomaly in the framework of the canonical formation of G2 stars, were brought to light \citep[see more details on both issues in review papers by][]{BastianLardo2018,Grattonetal19,MiloMari2022}. However, a detailed model analysis by \citet{bastianetal2015} showed that neither separate sources nor their combination are consistent with observations, particularly in terms of the helium abundance. Moreover, typical lithium abundance in G2 stars requires the dilution of the expelled enriched (but Li-free) gas with the pristine gas \citep{Prantzosetal2007}.

Conversely, we suggested \citep{Kravtsov2019,Kravtsov2020,Kravtsov_Calderon2021,Kravtsovetal2022,Dibetal2022,Kravtsovetal2024} an alternative approach for the origin of G2 stars due to the merger\slash collision of G1 low-mass MS stars, which is of particular relevance in the high-density GC environment\footnote{Note that we do not mean the merger\slash collision of massive stars, particularly in binaries, which is considered within the framework of canonical formation of G2 stars \citep[e.g.,][]{Sillsgleb2010,Wangetal2020}. Its incidence was probably high in the early GCs \citep{PortegiesZwartetal1999}. It is also argued \citep{deMinketal2014} to be notable among field stars.}. \citet{Kravtsovetal2022} showed some empirical dependencies suggesting this approach and estimated, in general terms, that it does not obviously suffer from the mass-budget problem. \citet{Kravtsovetal2024} looked into the matter in more detail. We demonstrated that a variety of data on close\slash hard binary systems and their evolution in different environments imply that it is primarily high-mass-ratio hard binaries of low-mass MS stars that are probably the main actors of the merger process mainly contributing to the formation of G2 stars. The fraction of G2 stars (1) depends primarily on both the fraction of such binaries merged and the slope of the initial mass function (IMF) in its most relevant low-mass part, and (2) can vary in a large range by varying realistic fractions of merged binaries even at a usual Milky Way-like IMF \citep{Kravtsovetal2024}. 

The formation of G2 stars by the merger of low-mass G1 MS stars implies that abundance anomalies of proton-capture elements (AAPCEL) in such G2 stars are also caused by the merger process. However, the issue of low-mass merger and its effect on AAPCEL in merger products\slash remnants is now not adopted for GCs \footnote{AAPCEL primarily refers to the abundance of CNO (and Na) as the most indicative, especially photometrically, for G2 stars in GCs.}. It is not only a poorly studied issue but also virtually omitted in the context of GC stellar populations because the origin of AAPCEL in G2 stars is unambiguously associated with other sources fully relevant for the canonical formation of G2 stars, with no other alternative. Perhaps for this reason, the ability of the merger process to cause AAPCEL in GC G2 stars is typically challenged. This leads to a paradoxical situation. Indeed, such a casual relationship between the merger of low-to-intermediate-mass stars and the arising of G2-like AAPCEL in the respective merger products is studied in the context of the origin of particular field stars (and phenomena) and of their elemental abundances. Surprising enough, this relationship is not virtually associated with GCs where the specific merger rate is believed to be significantly higher than that in the field. 

In this paper, we aim (1) to highlight the accumulated supporting evidence that the highly probable low-mass merger remnants in the field exhibit G2-like AAPCEL and (2) to show the signs of a specific channel tentatively identified as leading to the merger formation of the so-called extreme G2 RGB component. We argue that the impact of merger on the observed AAPCEL or its photometric effects cannot be negligible in GCs, and therefore at least a fraction of (RGB) stars isolated there as G2 stars exhibit AAPCEL caused by this process.

The present paper is organized in the following way. In the next section (Section~\ref{two_altern}), we show that presently there are two views on the origin of field stars with the G2-like abundance anomalies. In Section~\ref{results}, consisting of 5 Subsections, we describe the results of our overview and analysis of the published data on field stars with G2-like AAPCEL, as well as our proposal of a specific channel of the merger formation of G2 RGB stars. Summary and conclusions are in Section~\ref{summary}.

\section{Two views on field stars with the G2-like abundance anomalies}
\label{two_altern}

Based on the data available in the literature, we took care to select data about the most probable low-mass stars that (i) do not originate from GCs and (ii) exhibit G2-like AAPCEL, and thereby separate them from similar stars that, by their available characteristics, might either escape from the presently existing GCs or belonged earlier to disintegrated GCs. 
Reviewing the literature on low-mass stars with AAPCEL in the Galactic field, we finally achieved, along with the main goal, a fairly unexpected additional conclusion. Specifically, we have encountered conditionally two almost non-intersecting lines of investigations that both uncover field stars with G2-like AAPCEL but reach a different conclusion about its origin. 

One line aims at detecting stars in the Galactic field (in the bulge, disk, and halo) with G2-like AAPCEL, primarily with the CNO(Na) abundance anomalies, the origin of which one typically associates with loss from GCs. 
At least hundreds of red giant stars with G2-like AAPCEL and metallicities [Fe/H] typical for Galactic GCs were revealed in the Galactic field, mainly based on the APOGEE survey data. In particular, from these data gathered in the Galactic bulge, \citet{Schiavonetal2017} discovered a population of field stars with a high [N/Fe] ratio correlated with [Al/Fe] and anti-correlated with [C/Fe] ratios, typical for G2 stars, and a metallicity distribution function maximum around [Fe/H]$\approx-1.0$. \citet{FernandTrinetal2021} isolated even more metal-rich, on average, population of red giants, with G2-like AAPCEL and the [Fe/H] ratio extending to slightly sub-solar values. Moreover, \citet{Fernandezetal2017} discovered a small number of field red giants in the low metallicity range ($-1.8 <$ [Fe/H] $< -0.7$) strongly enriched in N, Na, and Al and depleted in C, O, and Mg. \citet{FernandTrinetal2022} isolated a much larger sample of 149 N-rich ([N/Fe] $\gtrsim +0.5$) field red giant stars throughout the Galaxy (i.e., toward the bulge, metal-poor disk and halo). Their G2-like abundance patterns show not only enrichment in the [N/Fe] ratio, but also depletion in the [C/Fe] ratio ([C/Fe] $< +0.15$). These stars also cover a wide range of metallicities ($-1.8 <$ [Fe/H] $< -0.7$) typical for GCs. The interested reader is referred to Table 2 from \citet{FernandTrinetal2022} for a summary of the information on the populations of the carbon-depleted nitrogen-enriched stars revealed in the field and on the respective publications. 

The authors of the aforementioned results argue that the field stars with G2-like AAPCEL have been stripped from existing GCs or from destroyed (fully or partially) GCs \citep{FernandTrinetal2021}, given the similarities in AAPCEL between these field stars and GC G2 stars, including the metallicity range in which they fall. \citet{FernandTrinetal2022} noted a diversity of kinematical and dynamical characteristics of the revealed N-rich field giants, which implies that they do not originate from the same birthplaces in the Galaxy. 
However, in strict terms, it has yet to be established how many of the revealed stars could previously belong to GCs. These findings cannot rule out the possibility that a fraction of these stars might have been originally formed in situ in the field but not in GCs (excluding, of course, the postulation that any star with G2-like AAPCEL unequivocally originated from a GC). In order to distinguish between these two alternatives, additional information on key stellar characteristics is required, such as reliably (asteroseismically) derived masses, luminosities, and evolutionary status, i.e. whether they are shell H-burning RGB stars or core He-burning red clump \slash HB stars.

Another line of research focuses on stars (and phenomena, too) highly probably formed in situ as field or open cluster constituents exhibiting mass \citep[e.g.,][]{Lietal2022} or\slash and abundance anomalies (i.e., G2-like AAPCEL) or peculiar locations in the color-magnitude diagram of open clusters \citep[e.g.,][]{Gelleretal2017a,Gelleretal2017b,Matteuzzietal2024}. The availability of large databases accumulated owing to advanced ground- and space-based surveys targeting spectroscopic, photometric, and astrometric characteristics of Galactic stars allows one to study deeper into various problems, including those of stellar astrophysics addressed here. Among them are the most feasible process(es) casing the mentioned anomalies. Specifically, interactions between stars in binary systems that lead to the formation of peculiar stars in the Galactic field and open clusters as a result of either mass transfer between the components or their mergers.  

We pay special attention to those stars with G2-like AAPCEL, the origin of which from GCs is improbable given their stellar characteristics.
We primarily refer to the so-called carbon-deficient (red clump) giants (see more details below), particularly to their low-mass fraction with $1.0 M_{\sun} < M \lesssim 2.0 M_{\sun}$, that is, expected to have passed through the helium flash at the end of the RGB, like (extended) horizontal branch (HB) stars in the present-day GC.

\section{The results}
\label{results}

We focus on published data accumulated on low-mass stars likely formed in situ in the field and exhibiting G2-like AAPCEL that is explained by mechanisms other than those proposed for G2 stars in GCs. So, we separated such stars from those excluded from our consideration because they were suggested to have escaped from GCs.

\subsection{Likely core helium-burning low-mass merger remnants in the field}
\label{field_mergers}

We primarily selected data published on stars with reliably estimated stellar mass from asteroseismology and with largely thin-disk kinematics. Although the [Fe/H] ratio is a less rigid discriminating parameter compared to the mass and kinematics, it is also quite useful in addition to them. We considered stars with metallicity biased towards solar and super-solar values. This further decreases the probability that such field stars with G2-like AAPCEL previously belonged to GCs. Despite the fact that the number of GCs known in the Milky Way has exceeded 200, none of them is known to have supersolar metallicity \citep{GarroMinnFernan2024}, and the fraction of GCs with slightly subsolar metallicity is low.

We emphasize on stars that in terms of evolution are core-helium-burning stars with G2-like AAPCEL. These are red-clump stars and hot helium (He) subdwarfs (He-sdOB). Very probably, these did not originate from GCs. Indeed, (i) they are mainly metal-rich disk stars (especially He-sdOB which are of supersolar metallicity) and (ii) their masses fall in the low-mass range, but are (of red-clump stars) obviously higher than the typical present-day maximum stellar mass around the MS turnoff ($M_{TO} \lesssim 0.9 M_{\sun}$) and on the lower RGB in GCs, unless they are improbable descendants of the rare case of massive blue stragglers normally forming and residing near the centers of GCs, deep inside cluster potential wells. 

Such particular field stars are supporting evidence of the formation of G2-like stars in the field due to a mechanism(s) different from those proposed and believed to be responsible for the (canonical) formation of G2 stars in GCs.

\subsection{The abundance pattern and origin of carbon-deficient red giants in the field, }
\label{CDRG}

We primarily pay attention to the so-called carbon-deficient red giants (CDRGs) or "weak G-band stars", a small population of a peculiar class of stars kinematically mainly belonging to the Galactic disk, which are known to have been revealed long ago \citep[see, for example, a historical insight into the discovery and study of these stars in][]{Bond2019}. They were studied for many decades, increasing in their revealed number \citep[][and references therein]{Mabenetal2023a}, but until very recently they were believed to be subgiant branch (SGB) and RGB stars of intermediate mass, $\sim 2.5-5.0 M_{\sun}$. However, relying on a selected sample of such stars with the totality of available data (on photometry, asteroseismology, spectroscopy, and astrometry) \citet{Mabenetal2023b} reached a different conclusion for a fraction of such stars. Specifically, they found that, in contrast to previous estimates, almost all of the selected (with slightly subsolar mean metallicity) CDRGs are core helium-burning stars (i.e., red clump stars) of low mass ($\lesssim 2.0 M_{\sun}$), and of various luminosities. These authors reached the conclusion that the most likely origin of more luminous red clump stars (among the selected CDRGs) is the merger of a helium white dwarf (HeWD) with an RGB star. For normal luminosity red clump stars, one cannot distinguish between core He-flash pollution and lower-mass merger scenarios. 

The range and mean values of metallicity ([Fe/H]), CNO abundances, and asteroseismologically derived masses of a sample of 15 low-mass CDRGs in the Galactic field are shown in Table~\ref{CDRG1}. The data presented are based on and calculated from the data on the elemental abundances and masses (M$_{AVG}$) of the individual sample stars listed in \citet{Mabenetal2023b}. The values of M$_{AVG}$ were obtained by the authors by averaging three separate asteroseismic estimates of each star's mass using three different formulae.   

\begin{table*}
\caption{The range and mean values of metallicity, CNO abundances, and asteroseismologically derived masses of a sample of 15 low-mass carbon-deficient giants in the Galactic field}
\label{CDRG1}
\centering
\begin{tabular}{ccccc}
\hline \hline \noalign{\smallskip}
$\Delta$ [Fe/H] & $\Delta$ [C/Fe] & $\Delta$ [N/Fe] & $\Delta$ [O/Fe] & $\Delta$ M$_{AVG}$ (M$_{\sun}$)$^a$ \\

$\overline{[Fe/H]}$ & $\overline{[C/Fe]}$ & $\overline{[N/Fe]}$ & $\overline{[O/Fe]}$ & $\overline{M}_{AVG}$ (M$_{\sun}$) \\
\hline \noalign{\smallskip}
-0.74 -- 0.08 & -0.87 -- -0.41 & 0.47 -- 0.80 & -0.11 -- 0.49 & 1.19 -- 2.34 \\
-0.18 & -0.56 & 0.59 & 0.05 & 1.65 \\

 \hline
\end{tabular}\\

\footnotetext{a}{$^a$ The data presented are based on and calculated from the original data on the elemental abundances and masses of individual stars listed in \citet{Mabenetal2023b}.} \\ 

\end{table*}

We also point out to very recent results on CDRGc obtained by \citet{Holandetal2024}. From their comprehensive chemical analysis of four poorly studied CDRGs (HD 54627, HD 105783, HD 198718, and HD 201557), they derived indicative elemental abundances of these CDRGs, in particular, in CNO and Na. Furthermore, \citet{Holandetal2023} found that the CDRG HD 16424 exhibits an enrichment in N and a depletion in C of [N/Fe]$ = +0.97$ and [C/Fe]$ = -0.57$, respectively, typical of a CDRG. They also observed a Na overabundance, a high abundance of Li-7, and a low mass ($1.61 M_{\sun}$) of HD 16424. The kinematics and the absence of alpha-element enrichment ($-0.12 \leq$ [$\alpha$/Fe] $\leq 0.06$ dex) indicate that these are stars of the thin disk. In Table~\ref{CDRG2}, as in Table~\ref{CDRG1}, we give the range and mean values of metallcity ([Fe/H]), CNO and Na abundances, and masses (derived using a different method than the one based on asteroseismology) of the five aforementioned CDRGs studied by \citet{Holandetal2023,Holandetal2024}.

\begin{table*}
\caption{The range and mean values of metallcity, CNO and Na abundances, and masses of a sample of 5 carbon-deficient giants in the Galactic field}
\label{CDRG2}
\centering
\begin{tabular}{cccccc}
\hline 
\hline 
\noalign{\smallskip}
$\Delta$ [Fe/H] & $\Delta$ [C/Fe] & $\Delta$ [N/Fe] & $\Delta$ [O/Fe] & $\Delta$ [Na/Fe] & $\Delta$ M (M$_{\sun}$)$^a$ \\

$\overline{[Fe/H]}$ & $\overline{[C/Fe]}$ & $\overline{[N/Fe]}$ & $\overline{[O/Fe]}$ & $\overline{[Na/Fe]}$ & $\overline{M}$ (M$_{\sun}$) \\
\hline \noalign{\smallskip}
-0.38 -- +0.09 & -1.77 -- -0.57 & +0.95 -- +1.12 & -0.27 -- +0.34 & +0.29 -- +0.37 & 1.61 -- 3.08 \\
-0.08 & -1.32 & 1.00 & -0.04 & +0.33 & 2.56 \\

 \hline
\end{tabular}\\

\footnotetext{a}{$^a$ The data given are based on and calculated from the original data on the elemental abundances and masses of individual stars taken from \citet{Holandetal2023,Holandetal2024}.} \\ 

\end{table*}

\citet{Bond2019} showed that CDRGs lie at systematically larger distances from the Galactic plane than normal giants, possibly indicating a role of binary mass transfer and mergers. He also summarized (see his references to publications on CNO abundance analyses of CDRGs) that the surfaces of CDRGs "are strongly contaminated with material that was once deep in the hydrogen-burning core".

One may conclude \citep[see, for example,][and respective references therein]{Mabenetal2023b} that in the first approximation, in terms of the G2-like anomalies of CNO and Na abundances (by omitting the subject related to possible effects of metallicity and stellar mass), at least a fraction of CDRGs are virtually indistinguishable (within the uncertainty) from GC G2 stars or from the populations of giants isolated in the Galactic field, whose origin is attributed to GCs.

\subsection{He intermediate rich hot sub-dwarfs with G2-like abundance anomaly in the field}
\label{hotSDWs}

Hot subdwarf stars (sdOBs) are known to be low-mass core helium-burning stars like both horizontal branch (HB) or red clump stars, but of high temperature and the temperature's large range. Their thin atmospheres are typically composed of pure hydrogen because of, in particular, the so-called gravitational settling. A fraction of sdOBs are helium-rich ones believed to form by two mechanisms: (i) the merger of two helium (He) white dwarfs (WDs) or a HeWD with a low-mass COWD; and (ii) common-envelope (CE) or mass-transfer (MT) stage of evolution in binaries \citep[see, in particular,][and references therein]{heber2016,PhilipMonetal2024}. The least numerous group of hot subdwarfs, referred to as intermediate helium-rich hot subdwarfs (iHe-sdOB), demonstrates a mixture of H and He in their atmospheres. It has been suggested that these stars might be in a transition phase of their evolution from the He-flash towards the HB. According to estimates made by \citet{AranRojetal2024}, most of the sdBs currently observed should descend from low-mass progenitors with initial masses $< 1.5$ M$_{\sun}$.

The case of sdOBs is complicated and ambiguous for spectral analysis because their large gravity, strong magnetic field, and high temperature lead to a number of processes such as radiative levitation and gravitational settling that affect their surface elemental abundance and its reliable determination. For this reason, and taking their relatively low luminosity into account, the number of these stars with detailed and reliable information on their elemental abundances is limited. However, \citet{DorschLatourHeber2019} found that two iHe-sdOBs, HZ 44 and HD 127493, show a strong CNO cycle pattern, with N being notably enriched while C and O are depleted with respect to the solar values. Here are in parentheses the individual values of the abundances of Fe (0.18; 1.00), C (-1.06; -1.41), N (1.46; 1.41), O (-0.91; $<-1.67$) and Na (1.01; $< 1.68$) of HZ 44 and HD 127493, respectively, from \citet{DorschLatourHeber2019}, who concluded that the CNO abundance of these two hot subdworfs can be explained by hydrogen burning in the CNO cycle. They also noted that despite the impact of atmospheric diffusion processes in these stars, "it is not plausible to assume that diffusion creates an abundance pattern of C, N, O, and Ne, that mimics the nucleosynthesis pattern so well". \citet{DorschLatourHeber2019} referred to the results of \citet{Jefferyetal2017} on a third iHe-sdOBs, [CW83]$0825+15$ (UVO $0825+15$), the abundance pattern of which "is also similar to HZ44 and HD127493 in that the CNO-cycle pattern is evident".

Separately note also the results on detailed abundance analysis of the helium-rich sdO  EC $20187-4939$ reported by \citet{Scottetal2023}. They found that this hot subdwarf with mass estimated to be $\sim 0.44$ M$_{\sun}$ also exhibits high N and low C and O, which is also consistent with the material processed by the CNO cycle. Based on the appropriate model, \citet{Scottetal2023} interpret EC $20187-4939$ as a binary merger. However, because of the high abundance of He, both components of the binary should probably be HeWDs.

\subsection{Appropriate models and a specific channel of the merger formation of G2 RGB stars}  
\label{MS_WD_binaries}

\citet{Zhangetal2017} proposed and modeled the merger of a HeWDs with a low-mass MS star (HeWD+MS) and showed that this kind of mergers can lead to the formation of some classes of hot subdwarfs or HB stars. From the detailed analysis of these models, \citet{Zhangetal2023} argued that the HeWD+MS merger can be at least one formation channel of the so-called BLAPs, blue large-amplitude pulsators, hot low-mass stars showing fast pulsational variability.

In the previous subsections, we highlighted a variety of observational results obtained by different authors on low-mass field stars with G2-like AAPCEL in the Galactic disk, who interpreted these objects as probable mergers from appropriate modeling. On the whole, these results are important observational signs that argue in favor of the ability of low-mass mergers to result in the merger product's abundance anomaly like that of G2 stars in GCs. 

\begin{figure}
\centering
\includegraphics[angle=-90,width=12.3cm]{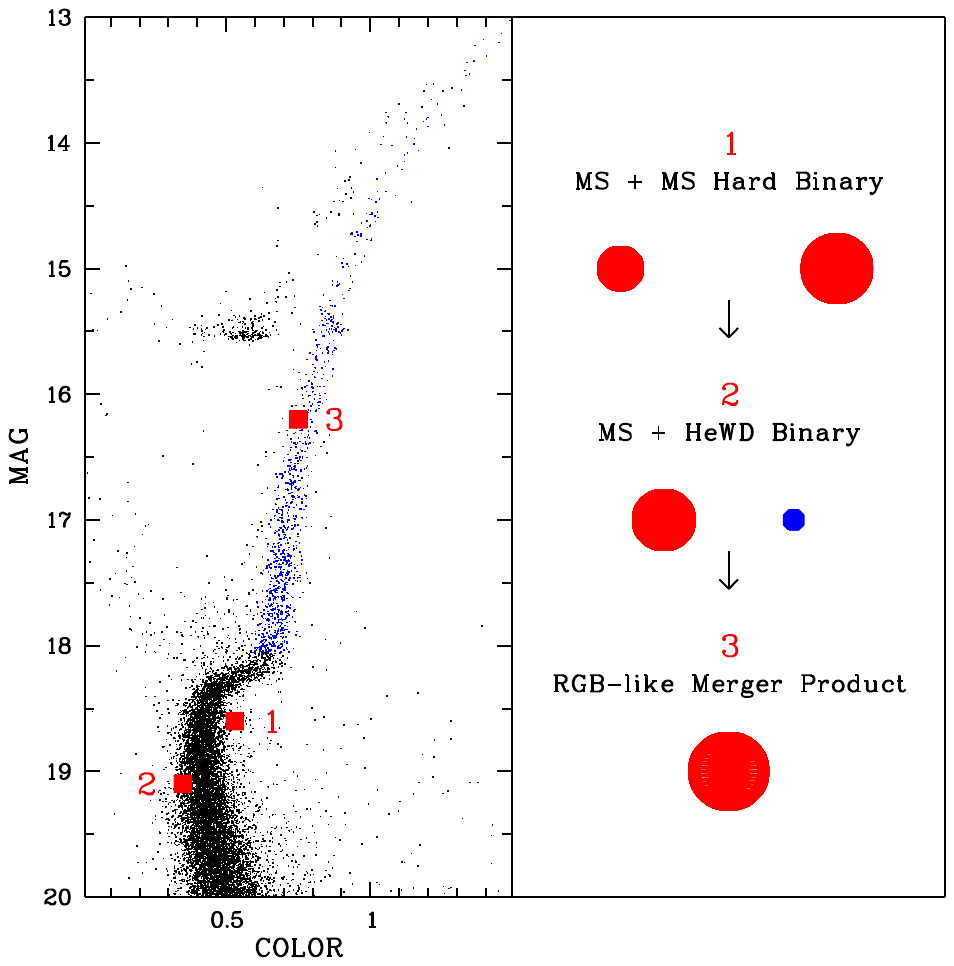}
\caption{Left panel: optical CMD, namely F555W-(F439W-F555W), of the globular cluster NGC 362 based on HST photometry from \citet{Piottoetal2002}; three red filled squires show the expected consecutive locations of an initial MS binary in its evolution toward RGB-like merger remnant, schematically shown in the right panel; the blue points show the stars of the RGB in its total range beginning from the RGB base. Right panel: schematic illustration of the main consecutive stages of the evolution of a hard binary initially composed of two MS srars (1), then converted into a (MSS$+$HeWD) binary (2), which finally merges forming RGB-like merger product (3) evolving along the cluster RGB.}
\label{CMD}
\end{figure}

At the same time, we have to note that the actual observations and available models allow us to suggest only a particular, specific channel of merger formation of a (minor) fraction of G2 stars observed on the RGB. It corresponds to the aforementioned models \citep{Zhangetal2017} of the mergers of low-mass HeWDs and MS stars with masses appropriate to those of the upper MS stars in GCs and uncovered by \citet{Dattatreyetal2023b} in NGC 362. This channel is schematically illustrated in Figure~\ref{CMD}.
In the high-density environment of GCs, the first transformation of an initial low-mass high-mass-ratio binary, composed of two upper MS stars (MSS$+$MSS) may result in a (MSS$+$HeWD) binary composed of a (extremely) low-mass HeWD and an upper MS star. This transformation is probably accompanied by the common-envelope (CE) stage of binary evolution in such a way that the mass of the secondary companion may increase insignificantly and it remains on the MS, without becoming a blue straggler. This outcome is possible, particularly in high-stellar environments. Moreover, \citet{parsonsetal2018} found that M dwarfs from a sample of close binaries with WDs appear to be indistinguishable from other M dwarfs, implying that the CE stage of evolution has a negligible impact on their structure. The aforementioned transformation is supported by observations that have revealed (MSS$+$HeWD) binaries in open\slash globular clusters, referred to as blue lurkers \citep{Leineretal2019}. They are typically located in the upper MS or around the MS turnoff in the optical color-magnitude diagram of star clusters \citep{ninetal2023,Dattatreyetal2023b,PanthiVaidya2024,jadhavetal2024}. A very important detail is that the cooling age of HeWDs estimated in such binaries in open clusters usually varies from a few tens of Myr to $\sim 1.0$ Gyr or even more, but is on average of the order of hundreds of Myr \citep{ninetal2023,Leineretal2025}. In contrast to open clusters, the cooling ages of extremely low-mass and low-mass HeWDs in similar binaries uncovered in NGC 362 \citep{Dattatreyetal2023b} are much younger: two are very young, within 0.1 Myr age, and the other have age between 1.8 and 4 Myr. Such a significant difference ($\sim 100$ times) between the timescales of (MSS$+$HeWD) binaries to reside in the low- and high-density stellar environments of open and GCs, respectively, is very demonstrative. Within our approach, this implies that the hard (MSS$+$HeWD) systems merge much faster in the GC environment than in the open cluster environment after their transformation from the (MSS$+$MSS) binaries. This is in agreement with the expected strong impact of the GC environment on the lifetime and fate of hard binaries. This stands as a preliminary conclusion, since the actual data on (MSS$+$HeWD) binaries are very limited, especially in GCs, and it is possible that the real situation is more complicated, with additional nuances. However, a useful estimate can be obtained.

\citet{Zhangetal2017} point that the products of low-mas (MSS$+$HeWD) mergers are expected to be RGB-like stars that likely evolve in a different way than normal RGB stars, including the RGB timescale. According to \citet{Zhangetal2017}, the evolution of a model merger product after the (MSS$+$HeWD) binary merger occurred and until the stage of a hot subdwarf or HB star takes a few tens of Myr. The mass of the interacting components that form the merger product that evolves to a hot subdwarf is, in particular, M$_{MSS}$$+$M$_{WD}$ = (0.65$+$0.30) M$_{\sun}$. Both M$_{MSS}$ and M$_{WD}$ may vary in a certain range. In the case of the evolution of the merger product to the HB, the tendency is that the proportion between the masses of the components changes compared to the evolution to a hot subdwarf in the sense that both M$_{MSS}$ itself and M$_{MSS}$\slash M$_{WD}$ ratio are a little higher at a decreasing M$_{WD}$ down to its limiting value of M$_{WD} =$0.25 M$_{\sun}$ among the calculated models. At this limit, the model value of M$_{MSS}$$+$M$_{WD}$ required to evolve to the HB falls within the range of values observed for the (MSS$+$HeWD) binaries found in NGC 362 \citep{Dattatreyetal2023b}, namely between 1.00 and 1.12 M$_{\sun}$, but at lower M$_{MSS}$ and higher M$_{WD}$. 

Taking the aforementioned difference between the characteristics of the model and observed (MSS$+$HeWD) binaries, the timescale of the RGB evolution of the RGB-like stars in NGC 362 may be even somewhat longer than expected for the model because of the lower mass of HeWDs in the (MSS$+$HeWD) binaries uncovered in NGC 362. Conservatively, we accepted this timescale to be 30 Myr. If we accept the mean timescale of the (MSS$+$HeWD) binaries to be around 3 Myr, then this implies that we may expect to find 10 times more the RGB-like merger products on the RGB of NGC 362 than the observed number of the (MSS$+$HeWD) binaries. We estimated the total actual number of RGB stars, from the RGB base to its tip (as shown in the left panel of Figure~\ref{CMD}), in NGC 362 as a whole to be around 1800 stars. This quantity has been obtained from the data on the population of RGB stars in NGC 362 from \citet{Kravtsov2009} who used HST photometry of \citet{Piottoetal2002} in the central parts of Galactic GCs. The fraction of RGB G2 stars in NGC 362 was obtained to be N$_{G2}$\slash N$_{TOT}$ = 0.72 using the fraction of G1 RGB stars from \citet{Milone_etal2017}. Finally, we accepted that the completeness of the (MSS$+$HeWD) binaries uncovered in NGC 362 (i.e., in the upper MS, around the MS turnoff) to be around 30\%. From here, the fraction of RGB-like stars, formed due to (MSS$+$HeWD) mergers, among G2 RGB stars in NGC 362 may be as high as 10\%. But it is rather an upper limit. 

We conclude that the specific channel [(MSS$+$MSS) binary -- (MSS$+$HeWD) binary -- RGB-like merger product] can contribute within $\sim$10\%, with uncertainty of the same order of magnitude, to G2 RGB stars in NGC 362. We tentatively identify it as the channel of the formation of the so-called extreme G2 RGB component isolated at a comparable fraction in GCs by \citep{Carretta2009}, which shows the largest deviation from G1 stars in the Na-O anti-correlation.

\subsection{Evidence of low-mass mergers of non-compact stars in the field}

We have earlier paid attention \citep{Kravtsovetal2022} to optical transients, the so-called luminous red novae (LRNe) occurring, in particular, in the Galactic field due to low (intermediate)-mass mergers, as being an important evidence of numerous mergers of low-mass stars in GCs over their long history in much more favorable conditions for stellar merger\slash collision in their dense central regions. This evidence became more comprehensive and insightful owing to new results, especially those on the chemical composition of LRNe remnants. In particular, \citet{Kaminsketal2023} found super-solar lithium abundances in a number of Galactic LRNe remnants thought to be caused by mergers involving non-compact low-mass stars such as (sub)giants and MS stars \citep[see also][and references therein]{kaminsky2024}. Also important is that, as in the case of the abundance anomalies observed in the high-probability field merger products\slash remnants considered in the previous subsections, \citet{tylendaetal2024} find ashes of hydrogen burning in the CNO cycles and also in the MgAl chain in the circumstellar gas of the remnant of the LRN CK Vul. Interesting enough, \citet{dorazigratton2020} reach the conclusion that spectroscopic investigations of both G1 and G2 stars converge towards the need for Li production due to the processes responsible for the occurrence of G2 stars in GCs.

Another probable type of merger manifestation, the ring-shaped ultraviolet nebula with a star at its center (TYC 2597-735-1), is reported by \citet{hoadleyetal2020}. From a detailed analysis of observations of a rare far-ultraviolet emitting object discovered earlier by the Galaxy Evolution Explorer, they argue that the best-fit models are those where the merger of old (i.e., of low mass) stars of the Galactic disk "happens after the primary begins evolving off its main sequence".

\section{Summary and conclusions}
\label{summary}

In the framework of our approach \citep{Kravtsovetal2022}, we tentatively identified \citep{Kravtsovetal2024} low-mass G1 MS hard binaries of high mass ratio as probable progenitors of G2 stars in GCs. It is known that these binaries merge much more efficiently and on a much shorter timescale in the GC environment than in the field or in open clusters. The merger of G1 low-mass MS binaries can overcome the so-called mass-budget problem, but its ability to result in the G2-like anomaly of light elements in low-mass stars-merger products are largely omitted. We discuss the evidence that such an outcome occurs by relying on highly probable merger remnants in the Galactic disk. In the following, we summarize our results and conclusions. 

We first recall that the mergers of low-mass stars occur even in the low-density environment of the Galactic field and open clusters despite its low efficiency. It was previously shown that the dynamical effects of tertiary companions in triple systems in the Galactic field lead to shorter periods (i.e., to increasing tightness) of the inner binaries compared to similar binaries without tertiary companions \citep{Tokovininetal2006,Tokovinin2023}. Currently, a growing body of evidence strengthens the understanding of the important role of hierarchical triple systems, particularly of low-mass (solar-like) stars, in the dynamical evolution of the inner binaries and in stimulating their final merger leading to the formation of rare peculiar stars, such as the B-type star LS V +22 25 with mass $ 1.1 \pm 0.5$ M$_{\sun}$ in the potential black hole binary LB-1 \citep{Irrgangetal2020}, and of compact objects up to a combination of triple WDs \citep[e.g.,][and references therein]{LagosVilchetal2024,Shariatetal2025,ArosBunsteretal2025,LiLuetal2025}. 
In addition to this effect on the tightness of the inner binaries in hierarchical triple systems, CE evolution is also an important process that makes the binaries closer or leads them to merge \citep[e.g.,][and references therein]{Ivanovaetal2013,PostnovYoung2014}, especially in the low-density environment of the Galactic field. In turn, the formation of some classes of hot subdwarfs in the field was proposed \citep{Zhangetal2017} to be due to the merger of a HeWD with a low-mass MS star, i.e. (HeWD$+$MS). 
It is relevant to note that the subject dealing with some fraction of Cepheids as possible inner binary mergers (in particular, between stars with individual masses below the Cepheids' lower mass limit) in hierarchical triple systems in the field and in open clusters is also currently being studied in different contexts \citep{Dinnbieretal2024,EspinozAranciPilec2025}.

The high-density stellar environment in GCs is especially favorable for the mergers of low-mass stars. This is due to at least two factors: (i) frequent close encounters between stars and their integral dynamical effect on binaries, including their hardening, and (2) the higher concentration of binary stars in the clusters' centers, implying that a high fraction of the binaries are involved in the process of their hardening and subsequent merger in systematically more dense, and therefore, more favorable environment. This process is expected to result in (much) higher specific rate of low-mass mergers in GCs compared to the field and open clusters. The most probable fate of the dynamical evolution of hard binaries in GCs is expected to be merger or collision, which has been argued long ago \citep{GoodmanHut1993,Fregeauetal2004}. In turn, \citet{MastrobuChurDav2021} reached the conclusion that collisions between pairs of MS stars are the most common stellar encounters leading to mergers, by modeling stellar encounters under the conditions relevant to the Galactic nuclear star cluster.

The mentioned dynamical effect of the tertiary companions in hierarchical triple systems on the systematically closer inner binaries compared to binaries with the same mass ratio without such companions strongly implies that initial binary stars in the dense environment of GCs should have formed systematically harder (closer) than in the field. Moreover, a larger initial fraction of low-mass binaries could have formed in GCs \citep{ivanovaetal2005}. These unique sites distinguish themselves not only by their high stellar density but also by their high star formation efficiency at their birth, similar to what is measured in massive clusters forming today in the Milky Way \citep{Dibetal2013}. In addition to the dynamical effects that make hard binaries harder in the dense environment of GCs, other effects might potentially favor the merger of binaries. For example, \citet{RoznerPerets2022} suggest that the properties of binaries in the early GCs could be affected by the gas produced by the evolved G1 stars. 

Owing to their unique characteristics, GCs are particular sites that are very favorable for the merger of (low-mass) stars. Therefore, notable populations of low-mass merger products (G2 stars) should have been formed and accumulated in GCs since their formation. This outcome and its effect are poorly considered in the context and relationship with the formation of G2 stars in GCs because the occurrence of G2-like abundance anomaly is unequivocally associated with other proposed sources within the framework of the canonical formation of G2 stars. Moreover, the ability of the merger of low-mass stars to cause AAPCEL in merger products seems to be neglected.

Here, we focus on this important issue. We point out to the increasing evidence that in the Galactic disk there are low ($1.0 M_{\sun} < M \lesssim 2.0 M_{\sun}$)-to-intermediate-mass red-clump giants with the thin disk kinematics, slightly subsolar mean metallicity, and G2-like abundance anomaly, the so-called carbon-deficient giants, which could hardly originate from GCs. Moreover, there are hot (He-intermediate rich) subdwarfs with a similar anomaly at a supersolar metallically. They were interpreted in the original papers as highly probable remnants of stellar mergers. There is also strong observational evidence of non-compact low-mass stellar mergers occurring in the Galactic field. This evidence strongly implies that GCs should contain a substantial number of such stars, in contrast to the widely accepted formation of G2 stars from gas enriched by the evolved G1 stars. Then, which objects can be identified with these stars in GCs? We argue that the totality of the evidence presented highlights the merger nature of at least a fraction of G2 stars in GCs. Both the available observations and models appropriate for GCs allow us to suggest a particular channel for the merger formation of a (minor) fraction of G2 stars observed on the RGB and more advanced evolutionary stages (or more exactly, at and above the MS turnoff). We estimate that this channel [(MSS$+$MSS) binary -- (MSS$+$HeWD) binary -- RGB-like merger product] can contribute within $\sim$10\% of G2 RGB stars. We tentatively identify it as the channel of the formation of the so-called extreme G2 RGB component isolated by \citet{Carretta2009} at a comparable fraction in Galactic GCs.

In this relation, we expect an anticorrelation [correlation] between the mean age of low-mass HeWDs in (MSS$+$HeWD) binaries and the encounter rate (mass) [the fraction of G1 RGB stars] of their parent GCs at a given metallicity.

The way G2 stars form in massive star clusters, in general, and in GCs, in particular, implies some expected effects that can be verified observationally. We here briefly review a number of them and compare their consistency with the formation mechanisms of G2 stars either via the canonical way or the merger of G1 low-mass MS binaries.
 
A fairly large piece of observational information about multiple populations detected outside of Galactic GCs, was obtained in the Magellanic Clouds' massive star clusters (MCMSCs). Convergent evidence has been accumulated that the onset of multiple populations in MCMSCs occurs at an age of about 2 Gyr \citep[e.g.,][]{Hollyheadetal2017,Niederhoferetal2017b,Hollyheadetal2018,martocchiaetal2018, Hollyheadetal2019,Milonetal2020,Lietal2021,martocchiaetal2021}. There is also indication that their onset in the MS can occur around 300--500 Myr earlier than in the RGB, that is, around the age of 1.7--1.5 Gyr \citep[e.g.,][]{Li2021,Cadelanoetal2022}. Moreover, a conclusion was reached about (i) a systematically higher fraction of G1 stars in MCMSCs of intermediate age than in GCs of the same present-day mass and (ii) an increasing range of light element abundance variation with increasing age of MCMSCs \citep[see][and references therein]{martocchiaetal2019,Salgadoetal2022}. Although these manifestations, apart from those noted in the Abstract, are very challenging for or even inconsistent with the canonical formation of G2 stars from the gas ejected from G1 polluters, they are in good agreement with a mechanism like the merger of MS G1 stars, which may essentially be extended in time and expected to initially manifest in the MS and then at more advanced stages of stellar evolution. The time delay of about 2 Gyr between the onset of multiple populations in MCMSCs and the cluster formation may be interpreted, in principle, in the framework of the merger paradigm as well.  For instance, the time delay between the occurrence of mergers \slash collisions and the beginning of changes in surface elemental abundances, particularly C (decreasing) and N (increasing) abundances of the merger\slash collision products of low-mass MS stars ($0.6M_{\sun}+0.6 M_{\sun}$) was deduced from simulations by \citet{Sillsetal2005} two decades ago. It is of the same order of magnitude, around 2 Gyr, and the elemental abundance changes increase for some time.   

The binary fraction among G1 and G2 stars are also important quantities and they provide a test to the merger-scenario of G2 stars. We have already noted in our recent work \citep{Kravtsovetal2024} that the merger formation of G2 stars implies that the G1 and G2 subpopulations in GCs should have notably different binary fractions in the sense of a much smaller fraction among G2 stars. This is in agreement with observations that showed that the binary fractions among G1 sub-populations in GCs are typically an order of magnitude higher than among G2 ones \citep[see][and references therein]{Grattonetal19}. Recent results obtained by \citet{Bortolanetal2025} and \citet{Milonetal2025} on the occurrence of binaries among G1 and G2 stars in quite distinct GCs, NGC 288 and 47 Tuc, confirmed the general trend found in earlier observations but with some details. Specifically, the essential difference between the binary fractions among the sub-populations is observed in the outer parts of the GCs, whereas it is much smaller in the cluster centers. The authors unambiguously interpret such a dissimilarity in the binary fractions due to distinct radial distributions of the sub-populations in the parent GCs, which result in differences in how G1 and G2 binaries are affected by dynamical effects. In contrast, we believe that the much lower fractions of binary stars among G2 subpopulations in GCs, in spite of intervening dynamical effects, are primarily the natural consequence of the merger origin of G2 stars. Recently obtained results by \citet{Muratoretal2024} on binary star fractions among slow and fast MS rotators in three Magellanic star clusters are consistent with our merger scenario. 

The manifestations of the multiple populations phenomenon in low- and very low-mass MS stars are also a valuable channel of observational information for understanding the nature of G2 stars.
Several results on G1 and G2 sub-populations on the MS at very low stellar masses were obtained in a series of papers based on HST photometry and published since the last decade. In particular, \citet{milonetal2012b,milonetal2014,milonetal2019,Dondoglioetal2022} studied multiple populations in low MS in several GCs, including the estimation of the fractions of G1 and G2 MS stars below the so-called MS knee (i.e., at $M_{MS} < 0.3 M_{\sun}$) but above $M_{MS} \sim 0.20 M_{\sun}$. Note here interesting differences in the fraction of G1 (G2) stars in the RGB and the lower MS. While in NGC 6752 it is reported to be the same within the error in both evolutionary sequences, in NGC 2808 and M4 it is different, especially in NGC 2808. In the latter GC, the fraction of G1 stars (characterized by \citet{milonetal2012b} as "primordial helium and enhanced carbon and oxygen abundances") in the mentioned mass range of the MS has been estimated to be around $\sim0.65$ (or G2 $\sim0.35$). Later, \citet{Dondoglioetal2022} showed that it varies radially in the cluster and decreases to $\sim 0.45$ (or G2 increases to $\sim0.55$) in the cluster center. However, it remains obviously larger than the fraction of G1 RGB stars, which was estimated to be around $0.232$ in \citet{Milone_etal2017}. In M4, the fractions of G1 stars in the same evolutionary sequences are (with the possible exception of the innermost region according to \citet{Dondoglioetal2022}) $\sim0.380$ \citep{milonetal2014} and $0.285$ \citep{Milone_etal2017}, respectively. Interestingly enough, these differences in the fraction of G1 stars between the MS and RGB in the same GCs are consistent with the result of the present paper on the specific merger channel contributing to the formation of RGB G2 stars. 

The process of merger of low-mass MS stars in a large range of stellar mass ($0.1 M_{\sun} < M_{MS} < 2.0 M_{\sun}$), and in a metallicity range relevant to the case of GCs, is poorly studied, particularly its impact on the surface abundance of light elements in the merger products. It is unclear how much typical oxygen enhancement is caused by such mergers in a GC and whether there is a dependency on the masses of the merged MS stars. In this context, the conclusions reached in their recent spectroscopic study by \citet{Marinoetal2024} that the difference in oxygen abundance between G1 and the most enhanced G2 stars in a sample of stars with masses in the range of $\sim0.4 - 0.5M_{\sun}$ is similar to that observed among RGB stars in the GC 47 Tuc, is consistent with the canonical formation of G2 stars. The consistency or inconsistency with their merger formation is unclear so far.  
                                                
Finally, we would like to point out to an intriguing detail. The demonstrative observational test for our approach would be the edge effect at the low-mass extreme of the MS in GCs and massive clusters of different mass, metallicity, age, etc. In its most general form (without taking various factors into account), it exploits the core idea of our scenario. In the framework of the merger origin of G2 stars down to the MS stars of the lowest mass, the low-mass limit of G2 MS stars should formally be at $M_{MS} \sim 0.16 M_{\sun}$, the duplicated mass of low-mass limit of G1 MS stars ($M_{min,MS} \approx 0.08 M_{\sun}$). At higher mass, the MS of a GC should be composed of both sub-populations with certain fractions \footnote{Qualitative formulation of our approach and respective simple analytical formalism for estimation of the fractions of G1 stars can be found in \citet{Kravtsovetal2024}.}. The absence of MS G2 stars is expected below the stellar mass of $M_{MS} \sim 0.16 M_{\sun}$. Therefore, the mass range of the MS composed only of G1 stars is $[0.08,0.16 M_{\sun}]$ at best. We refer to it as the "bottom MS mono-population tail". In reality, the tail may be shorter. The mass of the merger product of a twin binary system composed of two stars with the minimal MS mass may be somewhat smaller than $\sim 0.16 M_{\sun}$ due to the possible mass loss due to the merger (e.g., at the common envelope evolution stage) than the sum of the masses of the components. It is evident that the most appropriate test would be the most massive GCs whose deep gravitational wells can prevent the essential loss of stars of the lowest mass, that is, those of the G1 stars composing the bottom MS mono-population tail. A reliable verification of such a test is a challenging observational task, even with today's advanced facilities. Interestingly enough, using observations with JWST of 47 Tuc, one of the most massive Galactic GCs, \citet{marinmiletal2024} found a different low-mass extent of G1 and G2 subsequences of the cluster MS. Although the G2 subsequence is more populated than its G1 counterpart in the higher-mass parts of the MS, it becomes sparser towards its low-mass limit and apparently disappears at $M_{MS} \sim 0.10 M_{\sun}$.  \citet{marinmiletal2024} noted as the fact "the narrow sequence of ultracool stars", which "suggests a possible lack of very O-poor 2P stars among ultracool stars". In contrast, the G1 subsequence extends to its tentatively identified hydrogen burning limit at $M_{MS} \approx 0.075 M_{\sun}$, below which the likely brown dwarfs are present.

We believe that, taken together, the above-discussed body of various manifestations of multiple populations observed in both Galactic GCs and Magellanib Clouds' massive star clusters shows rather better consistence with the merger origin of G2 stars than with their canonical formation from enriched gas. However, many aspects of the merger scenario need to be studied, and numerous closely related particular problems and questions need to be answered to achieve a greater understanding of the nature and origin of G2 stars.

\begin{acknowledgements}

The study was carried out under the state assignment of Lomonosov Moscow State University.
The authors thank the anonymous referee for the expressed interest in the presented results and for the useful comments that improved the manuscript.

\end{acknowledgements}

\bibliographystyle{aa}
\bibliography{paper_aa}

\begin{thebibliography}{100}
\expandafter\ifx\csname natexlab\endcsname\relax\def\natexlab#1{#1}\fi

\bibitem[{{Arancibia-Rojas} {et~al.}(2024){Arancibia-Rojas}, {Zorotovic}, {Vu{\v{c}}kovi{\'c}}, {Bobrick}, {Vos}, \& {Piraino-Cerda}}]{AranRojetal2024}
{Arancibia-Rojas}, E., {Zorotovic}, M., {Vu{\v{c}}kovi{\'c}}, M., {et~al.} 2024, \mnras, 527, 11184

\bibitem[{{Aros-Bunster} {et~al.}(2025){Aros-Bunster}, {Schreiber}, {Toloza}, {Hernandez}, {Belloni}, {El-Badry}, {Vanderbosch}, {Lagos-Vilches}, {G{\"a}nsicke}, \& {Koester}}]{ArosBunsteretal2025}
{Aros-Bunster}, C., {Schreiber}, M.~R., {Toloza}, O., {et~al.} 2025, \aap, 693, L11

\bibitem[{{Bastian} {et~al.}(2015){Bastian}, {Cabrera-Ziri}, \& {Salaris}}]{bastianetal2015}
{Bastian}, N., {Cabrera-Ziri}, I., \& {Salaris}, M. 2015, \mnras, 449, 3333

\bibitem[{{Bastian} {et~al.}(2013){Bastian}, {Lamers}, {de Mink}, {Longmore}, {Goodwin}, \& {Gieles}}]{Bastianetal2013}
{Bastian}, N., {Lamers}, H.~J.~G.~L.~M., {de Mink}, S.~E., {et~al.} 2013, \mnras, 436, 2398

\bibitem[{{Bastian} \& {Lardo}(2018)}]{BastianLardo2018}
{Bastian}, N. \& {Lardo}, C. 2018, \araa, 56, 83

\bibitem[{{Bond}(2019)}]{Bond2019}
{Bond}, H.~E. 2019, \apj, 887, 12

\bibitem[{{Bortolan} {et~al.}(2025){Bortolan}, {Bruce}, {Milone}, {Vesperini}, {Dondoglio}, {Legnardi}, {Muratore}, {Ziliotto}, {Cordoni}, {Lagioia}, {Marino}, \& {Tailo}}]{Bortolanetal2025}
{Bortolan}, E., {Bruce}, J., {Milone}, A.~P., {et~al.} 2025, \aap, 696, A220

\bibitem[{{Bragaglia} {et~al.}(2017){Bragaglia}, {Carretta}, {D'Orazi}, {Sollima}, {Donati}, {Gratton}, \& {Lucatello}}]{Bragagliaetal2017}
{Bragaglia}, A., {Carretta}, E., {D'Orazi}, V., {et~al.} 2017, \aap, 607, A44

\bibitem[{{Cadelano} {et~al.}(2022){Cadelano}, {Dalessandro}, {Salaris}, {Bastian}, {Mucciarelli}, {Saracino}, {Martocchia}, \& {Cabrera-Ziri}}]{Cadelanoetal2022}
{Cadelano}, M., {Dalessandro}, E., {Salaris}, M., {et~al.} 2022, \apjl, 924, L2

\bibitem[{Carretta {et~al.}(2009)Carretta, Bragaglia, Gratton, Lucatello, Catanzaro, Leone, Bellazzini, Claudi, D'Orazi, Momany, Ortolani, Pancino, Piotto, Recio-Blanco, \& Sabbi}]{Carretta2009}
Carretta, E., Bragaglia, A., Gratton, R.~G., {et~al.} 2009, Astronomy {\&} Astrophysics, 505, 117

\bibitem[{{Carretta} {et~al.}(2010){Carretta}, {Bragaglia}, {Gratton}, {Recio-Blanco}, {Lucatello}, {D'Orazi}, \& {Cassisi}}]{Carrettaetal2010a}
{Carretta}, E., {Bragaglia}, A., {Gratton}, R.~G., {et~al.} 2010, \aap, 516, A55

\bibitem[{{Dattatrey} {et~al.}(2023){Dattatrey}, {Yadav}, {Kumawat}, {Rani}, {Singh}, {Subramaniam}, \& {Singh}}]{Dattatreyetal2023b}
{Dattatrey}, A.~K., {Yadav}, R.~K.~S., {Kumawat}, G., {et~al.} 2023, \mnras, 523, L58

\bibitem[{{de Mink} {et~al.}(2014){de Mink}, {Sana}, {Langer}, {Izzard}, \& {Schneider}}]{deMinketal2014}
{de Mink}, S.~E., {Sana}, H., {Langer}, N., {Izzard}, R.~G., \& {Schneider}, F.~R.~N. 2014, \apj, 782, 7

\bibitem[{{Dib} {et~al.}(2013){Dib}, {Gutkin}, {Brandner}, \& {Basu}}]{Dibetal2013}
{Dib}, S., {Gutkin}, J., {Brandner}, W., \& {Basu}, S. 2013, \mnras, 436, 3727

\bibitem[{{Dib} {et~al.}(2022){Dib}, {Kravtsov}, {Haghi}, {Zonoozi}, \& {Belinch{\'o}n}}]{Dibetal2022}
{Dib}, S., {Kravtsov}, V.~V., {Haghi}, H., {Zonoozi}, A.~H., \& {Belinch{\'o}n}, J.~A. 2022, \aap, 664, A145

\bibitem[{{Dinnbier} {et~al.}(2024){Dinnbier}, {Anderson}, \& {Kroupa}}]{Dinnbieretal2024}
{Dinnbier}, F., {Anderson}, R.~I., \& {Kroupa}, P. 2024, \aap, 690, A385

\bibitem[{{Dondoglio} {et~al.}(2022){Dondoglio}, {Milone}, {Renzini}, {Vesperini}, {Lagioia}, {Marino}, {Bellini}, {Carlos}, {Cordoni}, {Jang}, {Legnardi}, {Libralato}, {Mohandasan}, {D'Antona}, {Martorano}, {Muratore}, \& {Tailo}}]{Dondoglioetal2022}
{Dondoglio}, E., {Milone}, A.~P., {Renzini}, A., {et~al.} 2022, \apj, 927, 207

\bibitem[{{D'Orazi} \& {Gratton}(2020)}]{dorazigratton2020}
{D'Orazi}, V. \& {Gratton}, R. 2020, \memsai, 91, 98

\bibitem[{{Dorsch} {et~al.}(2019){Dorsch}, {Latour}, \& {Heber}}]{DorschLatourHeber2019}
{Dorsch}, M., {Latour}, M., \& {Heber}, U. 2019, \aap, 630, A130

\bibitem[{{Espinoza-Arancibia} \& {Pilecki}(2025)}]{EspinozAranciPilec2025}
{Espinoza-Arancibia}, F. \& {Pilecki}, B. 2025, \apjl, 981, L35

\bibitem[{{Fern{\'a}ndez-Trincado} {et~al.}(2022){Fern{\'a}ndez-Trincado}, {Beers}, {Barbuy}, {Minniti}, {Chiappini}, {Garro}, {Tang}, {Alves-Brito}, {Villanova}, {Geisler}, {Lane}, \& {Diaz}}]{FernandTrinetal2022}
{Fern{\'a}ndez-Trincado}, J.~G., {Beers}, T.~C., {Barbuy}, B., {et~al.} 2022, \aap, 663, A126

\bibitem[{{Fern{\'a}ndez-Trincado} {et~al.}(2021){Fern{\'a}ndez-Trincado}, {Beers}, {Queiroz}, {Chiappini}, {Minniti}, {Barbuy}, {Majewski}, {Ortigoza-Urdaneta}, {Moni Bidin}, {Robin}, {Moreno}, {Chaves-Velasquez}, {Villanova}, {Lane}, {Pan}, \& {Bizyaev}}]{FernandTrinetal2021}
{Fern{\'a}ndez-Trincado}, J.~G., {Beers}, T.~C., {Queiroz}, A. B.~A., {et~al.} 2021, \apjl, 918, L37

\bibitem[{{Fern{\'a}ndez-Trincado} {et~al.}(2017){Fern{\'a}ndez-Trincado}, {Zamora}, {Garc{\'\i}a-Hern{\'a}ndez}, {Souto}, {Dell'Agli}, {Schiavon}, {Geisler}, {Tang}, {Villanova}, {Hasselquist}, {Mennickent}, {Cunha}, {Shetrone}, {Allende Prieto}, {Vieira}, {Zasowski}, {Sobeck}, {Hayes}, {Majewski}, {Placco}, {Beers}, {Schleicher}, {Robin}, {M{\'e}sz{\'a}ros}, {Masseron}, {Garc{\'\i}a P{\'e}rez}, {Anders}, {Meza}, {Alves-Brito}, {Carrera}, {Minniti}, {Lane}, {Fern{\'a}ndez-Alvar}, {Moreno}, {Pichardo}, {P{\'e}rez-Villegas}, {Schultheis}, {Roman-Lopes}, {Fuentes}, {Nitschelm}, {Harding}, {Bizyaev}, {Pan}, {Oravetz}, {Simmons}, {Ivans}, {Blanco-Cuaresma}, {Hern{\'a}ndez}, {Alonso-Garc{\'\i}a}, {Valenzuela}, \& {Chanam{\'e}}}]{Fernandezetal2017}
{Fern{\'a}ndez-Trincado}, J.~G., {Zamora}, O., {Garc{\'\i}a-Hern{\'a}ndez}, D.~A., {et~al.} 2017, \apjl, 846, L2

\bibitem[{{Fregeau} {et~al.}(2004){Fregeau}, {Cheung}, {Portegies Zwart}, \& {Rasio}}]{Fregeauetal2004}
{Fregeau}, J.~M., {Cheung}, P., {Portegies Zwart}, S.~F., \& {Rasio}, F.~A. 2004, \mnras, 352, 1

\bibitem[{{Garro} {et~al.}(2024){Garro}, {Minniti}, \& {Fern{\'a}ndez-Trincado}}]{GarroMinnFernan2024}
{Garro}, E.~R., {Minniti}, D., \& {Fern{\'a}ndez-Trincado}, J.~G. 2024, \aap, 687, A214

\bibitem[{{Geller} {et~al.}(2017{\natexlab{a}}){Geller}, {Leiner}, {Bellini}, {Gleisinger}, {Haggard}, {Kamann}, {Leigh}, {Mathieu}, {Sills}, {Watkins}, \& {Zurek}}]{Gelleretal2017a}
{Geller}, A.~M., {Leiner}, E.~M., {Bellini}, A., {et~al.} 2017{\natexlab{a}}, \apj, 840, 66

\bibitem[{{Geller} {et~al.}(2017{\natexlab{b}}){Geller}, {Leiner}, {Chatterjee}, {Leigh}, {Mathieu}, \& {Sills}}]{Gelleretal2017b}
{Geller}, A.~M., {Leiner}, E.~M., {Chatterjee}, S., {et~al.} 2017{\natexlab{b}}, \apj, 842, 1

\bibitem[{{Goodman} \& {Hut}(1993)}]{GoodmanHut1993}
{Goodman}, J. \& {Hut}, P. 1993, \apj, 403, 271

\bibitem[{{Gratton} {et~al.}(2019){Gratton}, {Bragaglia}, {Carretta}, {D'Orazi}, {Lucatello}, \& {Sollima}}]{Grattonetal19}
{Gratton}, R., {Bragaglia}, A., {Carretta}, E., {et~al.} 2019, \aapr, 27, 8

\bibitem[{{Heber}(2016)}]{heber2016}
{Heber}, U. 2016, \pasp, 128, 082001

\bibitem[{{Hoadley} {et~al.}(2020){Hoadley}, {Martin}, {Metzger}, {Seibert}, {McWilliam}, {Shen}, {Neill}, {Stefansson}, {Monson}, \& {Schaefer}}]{hoadleyetal2020}
{Hoadley}, K., {Martin}, D.~C., {Metzger}, B.~D., {et~al.} 2020, \nat, 587, 387

\bibitem[{{Holanda} {et~al.}(2023){Holanda}, {Drake}, \& {Pereira}}]{Holandetal2023}
{Holanda}, N., {Drake}, N.~A., \& {Pereira}, C.~B. 2023, \mnras, 518, 4038

\bibitem[{{Holanda} {et~al.}(2024){Holanda}, {Flaulhabe}, {Quispe-Huaynasi}, {Sonally}, \& {Pereira}}]{Holandetal2024}
{Holanda}, N., {Flaulhabe}, T., {Quispe-Huaynasi}, F., {Sonally}, A., \& {Pereira}, C.~B. 2024, \apj, 971, 152

\bibitem[{{Hollyhead} {et~al.}(2017){Hollyhead}, {Kacharov}, {Lardo}, {Bastian}, {Hilker}, {Rejkuba}, {Koch}, {Grebel}, \& {Georgiev}}]{Hollyheadetal2017}
{Hollyhead}, K., {Kacharov}, N., {Lardo}, C., {et~al.} 2017, \mnras, 465, L39

\bibitem[{{Hollyhead} {et~al.}(2018){Hollyhead}, {Lardo}, {Kacharov}, {Bastian}, {Hilker}, {Rejkuba}, {Koch}, {Grebel}, \& {Georgiev}}]{Hollyheadetal2018}
{Hollyhead}, K., {Lardo}, C., {Kacharov}, N., {et~al.} 2018, \mnras, 476, 114

\bibitem[{{Hollyhead} {et~al.}(2019){Hollyhead}, {Martocchia}, {Lardo}, {Bastian}, {Kacharov}, {Niederhofer}, {Cabrera-Ziri}, {Dalessandro}, {Mucciarelli}, {Salaris}, \& {Usher}}]{Hollyheadetal2019}
{Hollyhead}, K., {Martocchia}, S., {Lardo}, C., {et~al.} 2019, \mnras, 484, 4718

\bibitem[{{Irrgang} {et~al.}(2020){Irrgang}, {Geier}, {Kreuzer}, {Pelisoli}, \& {Heber}}]{Irrgangetal2020}
{Irrgang}, A., {Geier}, S., {Kreuzer}, S., {Pelisoli}, I., \& {Heber}, U. 2020, \aap, 633, L5

\bibitem[{{Ivanova} {et~al.}(2005){Ivanova}, {Belczynski}, {Fregeau}, \& {Rasio}}]{ivanovaetal2005}
{Ivanova}, N., {Belczynski}, K., {Fregeau}, J.~M., \& {Rasio}, F.~A. 2005, \mnras, 358, 572

\bibitem[{{Ivanova} {et~al.}(2013){Ivanova}, {Justham}, {Chen}, {De Marco}, {Fryer}, {Gaburov}, {Ge}, {Glebbeek}, {Han}, {Li}, {Lu}, {Marsh}, {Podsiadlowski}, {Potter}, {Soker}, {Taam}, {Tauris}, {van den Heuvel}, \& {Webbink}}]{Ivanovaetal2013}
{Ivanova}, N., {Justham}, S., {Chen}, X., {et~al.} 2013, \aapr, 21, 59

\bibitem[{{Jadhav} {et~al.}(2024){Jadhav}, {Subramaniam}, \& {Sagar}}]{jadhavetal2024}
{Jadhav}, V.~V., {Subramaniam}, A., \& {Sagar}, R. 2024, \aap, 688, A152

\bibitem[{{Jeffery} {et~al.}(2017){Jeffery}, {Baran}, {Behara}, {Kvammen}, {Martin}, {Naslim}, {{\O}stensen}, {Preece}, {Reed}, {Telting}, \& {Woolf}}]{Jefferyetal2017}
{Jeffery}, C.~S., {Baran}, A.~S., {Behara}, N.~T., {et~al.} 2017, \mnras, 465, 3101

\bibitem[{{Kaminski}(2024)}]{kaminsky2024}
{Kaminski}, T. 2024, arXiv e-prints, arXiv:2401.03919

\bibitem[{{Kami{\'n}ski} {et~al.}(2023){Kami{\'n}ski}, {Schmidt}, {Hajduk}, {Kiljan}, {Izviekova}, \& {Frankowski}}]{Kaminsketal2023}
{Kami{\'n}ski}, T., {Schmidt}, M., {Hajduk}, M., {et~al.} 2023, \aap, 672, A196

\bibitem[{{Kraft}(1979)}]{Kraft1979}
{Kraft}, R.~P. 1979, \araa, 17, 309

\bibitem[{{Kravtsov}(2019)}]{Kravtsov2019}
{Kravtsov}, V. 2019, Boletin de la Asociacion Argentina de Astronomia La Plata Argentina, 61, 122

\bibitem[{{Kravtsov} \& {Calder{\'o}n}(2021)}]{Kravtsov_Calderon2021}
{Kravtsov}, V. \& {Calder{\'o}n}, F.~A. 2021, \aj, 161, 7

\bibitem[{{Kravtsov} {et~al.}(2024){Kravtsov}, {Dib}, \& {Calder{\'o}n}}]{Kravtsovetal2024}
{Kravtsov}, V., {Dib}, S., \& {Calder{\'o}n}, F.~A. 2024, \mnras, 527, 7005

\bibitem[{{Kravtsov} {et~al.}(2022){Kravtsov}, {Dib}, {Calder{\'o}n}, \& {Belinch{\'o}n}}]{Kravtsovetal2022}
{Kravtsov}, V., {Dib}, S., {Calder{\'o}n}, F.~A., \& {Belinch{\'o}n}, J.~A. 2022, \mnras, 512, 2936

\bibitem[{{Kravtsov}(2009)}]{Kravtsov2009}
{Kravtsov}, V.~V. 2009, \aj, 137, 5110

\bibitem[{{Kravtsov}(2020)}]{Kravtsov2020}
{Kravtsov}, V.~V. 2020, Boletin de la Asociacion Argentina de Astronomia La Plata Argentina, 61C, 57

\bibitem[{{Lagos-Vilches} {et~al.}(2024){Lagos-Vilches}, {Hernandez}, {Schreiber}, {Parsons}, \& {G{\"a}nsicke}}]{LagosVilchetal2024}
{Lagos-Vilches}, F., {Hernandez}, M., {Schreiber}, M.~R., {Parsons}, S.~G., \& {G{\"a}nsicke}, B.~T. 2024, \mnras, 534, 3229

\bibitem[{{Leiner} {et~al.}(2019){Leiner}, {Mathieu}, {Vanderburg}, {Gosnell}, \& {Smith}}]{Leineretal2019}
{Leiner}, E., {Mathieu}, R.~D., {Vanderburg}, A., {Gosnell}, N.~M., \& {Smith}, J.~C. 2019, \apj, 881, 47

\bibitem[{{Leiner} {et~al.}(2025){Leiner}, {Gosnell}, {Geller}, {Sun}, {Mathieu}, \& {Sills}}]{Leineretal2025}
{Leiner}, E.~M., {Gosnell}, N.~M., {Geller}, A.~M., {et~al.} 2025, \apjl, 979, L1

\bibitem[{{Li}(2021)}]{Li2021}
{Li}, C. 2021, \apj, 921, 171

\bibitem[{{Li} {et~al.}(2021){Li}, {Tang}, {Milone}, {de Grijs}, {Hong}, {Yang}, \& {Wang}}]{Lietal2021}
{Li}, C., {Tang}, B., {Milone}, A.~P., {et~al.} 2021, \apj, 906, 133

\bibitem[{{Li} {et~al.}(2022){Li}, {Bedding}, {Murphy}, {Stello}, {Chen}, {Huber}, {Joyce}, {Marks}, {Zhang}, {Bi}, {Colman}, {Hayden}, {Hey}, {Li}, {Montet}, {Sharma}, \& {Wu}}]{Lietal2022}
{Li}, Y., {Bedding}, T.~R., {Murphy}, S.~J., {et~al.} 2022, Nature Astronomy, 6, 673

\bibitem[{{Li} {et~al.}(2025){Li}, {Lu}, {L{\"u}}, {Zhu}, {Liu}, \& {Yu}}]{LiLuetal2025}
{Li}, Z., {Lu}, X., {L{\"u}}, G., {et~al.} 2025, \apjl, 979, L37

\bibitem[{{Maben} {et~al.}(2023{\natexlab{a}}){Maben}, {Campbell}, {Kumar}, {Reddy}, \& {Zhao}}]{Mabenetal2023b}
{Maben}, S., {Campbell}, S.~W., {Kumar}, Y.~B., {Reddy}, B.~E., \& {Zhao}, G. 2023{\natexlab{a}}, \apj, 957, 18

\bibitem[{{Maben} {et~al.}(2023{\natexlab{b}}){Maben}, {Kumar}, {Reddy}, {Campbell}, \& {Zhao}}]{Mabenetal2023a}
{Maben}, S., {Kumar}, Y.~B., {Reddy}, B.~E., {Campbell}, S.~W., \& {Zhao}, G. 2023{\natexlab{b}}, \mnras, 525, 4554

\bibitem[{{Marino} {et~al.}(2024{\natexlab{a}}){Marino}, {Milone}, {Legnardi}, {Renzini}, {Dondoglio}, {Cavecchi}, {Cordoni}, {Dotter}, {Lagioia}, {Ziliotto}, {Bernizzoni}, {Bortolan}, {Carlos}, {Jang}, {Mohandasan}, {Muratore}, \& {Tailo}}]{marinmiletal2024}
{Marino}, A.~F., {Milone}, A.~P., {Legnardi}, M.~V., {et~al.} 2024{\natexlab{a}}, \apj, 965, 189

\bibitem[{{Marino} {et~al.}(2024{\natexlab{b}}){Marino}, {Milone}, {Renzini}, {Dondoglio}, {Bortolan}, {Carlos}, {Cordoni}, {Dotter}, {Jang}, {Lagioia}, {Legnardi}, {Muratore}, {Mohandasan}, {Tailo}, \& {Ziliotto}}]{Marinoetal2024}
{Marino}, A.~F., {Milone}, A.~P., {Renzini}, A., {et~al.} 2024{\natexlab{b}}, \apjl, 969, L8

\bibitem[{{Martocchia} {et~al.}(2019){Martocchia}, {Dalessandro}, {Lardo}, {Cabrera-Ziri}, {Bastian}, {Kozhurina-Platais}, {Salaris}, {Chantereau}, {Geisler}, {Hilker}, {Kacharov}, {Larsen}, {Mucciarelli}, {Niederhofer}, {Platais}, \& {Usher}}]{martocchiaetal2019}
{Martocchia}, S., {Dalessandro}, E., {Lardo}, C., {et~al.} 2019, \mnras, 487, 5324

\bibitem[{{Martocchia} {et~al.}(2021){Martocchia}, {Lardo}, {Rejkuba}, {Kamann}, {Bastian}, {Larsen}, {Cabrera-Ziri}, {Chantereau}, {Dalessandro}, {Kacharov}, \& {Salaris}}]{martocchiaetal2021}
{Martocchia}, S., {Lardo}, C., {Rejkuba}, M., {et~al.} 2021, \mnras, 505, 5389

\bibitem[{{Martocchia} {et~al.}(2018){Martocchia}, {Niederhofer}, {Dalessandro}, {Bastian}, {Kacharov}, {Usher}, {Cabrera-Ziri}, {Lardo}, {Cassisi}, {Geisler}, {Hilker}, {Hollyhead}, {Kozhurina-Platais}, {Larsen}, {Mackey}, {Mucciarelli}, {Platais}, \& {Salaris}}]{martocchiaetal2018}
{Martocchia}, S., {Niederhofer}, F., {Dalessandro}, E., {et~al.} 2018, \mnras, 477, 4696

\bibitem[{{Mastrobuono-Battisti} {et~al.}(2021){Mastrobuono-Battisti}, {Church}, \& {Davies}}]{MastrobuChurDav2021}
{Mastrobuono-Battisti}, A., {Church}, R.~P., \& {Davies}, M.~B. 2021, \mnras, 505, 3314

\bibitem[{{Matteuzzi} {et~al.}(2024){Matteuzzi}, {Hendriks}, {Izzard}, {Miglio}, {Brogaard}, {Montalb{\'a}n}, {Tailo}, \& {Mazzi}}]{Matteuzzietal2024}
{Matteuzzi}, M., {Hendriks}, D., {Izzard}, R.~G., {et~al.} 2024, \aap, 691, A17

\bibitem[{{Milone} \& {Marino}(2022)}]{MiloMari2022}
{Milone}, A.~P. \& {Marino}, A.~F. 2022, Universe, 8, 359

\bibitem[{{Milone} {et~al.}(2019){Milone}, {Marino}, {Bedin}, {Anderson}, {Apai}, {Bellini}, {Dieball}, {Salaris}, {Libralato}, {Nardiello}, {Bergeron}, {Burgasser}, {Rees}, {Rich}, \& {Richer}}]{milonetal2019}
{Milone}, A.~P., {Marino}, A.~F., {Bedin}, L.~R., {et~al.} 2019, \mnras, 484, 4046

\bibitem[{{Milone} {et~al.}(2014){Milone}, {Marino}, {Bedin}, {Piotto}, {Cassisi}, {Dieball}, {Anderson}, {Jerjen}, {Asplund}, {Bellini}, {Brogaard}, {Dotter}, {Giersz}, {Heggie}, {Knigge}, {Rich}, {van den Berg}, \& {Buonanno}}]{milonetal2014}
{Milone}, A.~P., {Marino}, A.~F., {Bedin}, L.~R., {et~al.} 2014, \mnras, 439, 1588

\bibitem[{{Milone} {et~al.}(2025){Milone}, {Marino}, {Bernizzoni}, {Muratore}, {Legnardi}, {Barbieri}, {Bortolan}, {Bouras}, {Bruce}, {Cordoni}, {D'Antona}, {Dell'Agli}, {Dondoglio}, {Grimaldi}, {Jang}, {Lagioia}, {Lee}, {Lionetto}, {Mohandasan}, {Pang}, {Pianta}, {Posenato}, {Renzini}, {Tailo}, {Ventura}, {Ventura}, {Vesperini}, \& {Ziliotto}}]{Milonetal2025}
{Milone}, A.~P., {Marino}, A.~F., {Bernizzoni}, M., {et~al.} 2025, \aap, 698, A247

\bibitem[{{Milone} {et~al.}(2012){Milone}, {Marino}, {Cassisi}, {Piotto}, {Bedin}, {Anderson}, {Allard}, {Aparicio}, {Bellini}, {Buonanno}, {Monelli}, \& {Pietrinferni}}]{milonetal2012b}
{Milone}, A.~P., {Marino}, A.~F., {Cassisi}, S., {et~al.} 2012, \apjl, 754, L34

\bibitem[{{Milone} {et~al.}(2020){Milone}, {Marino}, {Da Costa}, {Lagioia}, {D'Antona}, {Goudfrooij}, {Jerjen}, {Massari}, {Renzini}, {Yong}, {Baumgardt}, {Cordoni}, {Dondoglio}, {Li}, {Tailo}, {Asa'd}, \& {Ventura}}]{Milonetal2020}
{Milone}, A.~P., {Marino}, A.~F., {Da Costa}, G.~S., {et~al.} 2020, \mnras, 491, 515

\bibitem[{{Milone} {et~al.}(2017){Milone}, {Piotto}, {Renzini}, {Marino}, {Bedin}, {Vesperini}, {D'Antona}, {Nardiello}, {Anderson}, {King}, {Yong}, {Bellini}, {Aparicio}, {Barbuy}, {Brown}, {Cassisi}, {Ortolani}, {Salaris}, {Sarajedini}, \& {van der Marel}}]{Milone_etal2017}
{Milone}, A.~P., {Piotto}, G., {Renzini}, A., {et~al.} 2017, \mnras, 464, 3636

\bibitem[{{Muratore} {et~al.}(2024){Muratore}, {Milone}, {D'Antona}, {Nastasio}, {Cordoni}, {Legnardi}, {He}, {Ziliotto}, {Dondoglio}, {Bernizzoni}, {Tailo}, {Bortolan}, {Dell'Agli}, {Deng}, {Lagioia}, {Li}, {Marino}, \& {Ventura}}]{Muratoretal2024}
{Muratore}, F., {Milone}, A.~P., {D'Antona}, F., {et~al.} 2024, \aap, 692, A135

\bibitem[{{Nataf} {et~al.}(2019){Nataf}, {Wyse}, {Schiavon}, {Ting}, {Minniti}, {Cohen}, {Fern{\'a}ndez-Trincado}, {Geisler}, {Nitschelm}, \& {Frinchaboy}}]{Nataf_etal2019}
{Nataf}, D.~M., {Wyse}, R. F.~G., {Schiavon}, R.~P., {et~al.} 2019, \aj, 158, 14

\bibitem[{{Niederhofer} {et~al.}(2017){Niederhofer}, {Bastian}, {Kozhurina-Platais}, {Larsen}, {Hollyhead}, {Lardo}, {Cabrera-Ziri}, {Kacharov}, {Platais}, {Salaris}, {Cordero}, {Dalessandro}, {Geisler}, {Hilker}, {Li}, {Mackey}, \& {Mucciarelli}}]{Niederhoferetal2017b}
{Niederhofer}, F., {Bastian}, N., {Kozhurina-Platais}, V., {et~al.} 2017, \mnras, 465, 4159

\bibitem[{{Nine} {et~al.}(2023){Nine}, {Mathieu}, {Gosnell}, \& {Leiner}}]{ninetal2023}
{Nine}, A.~C., {Mathieu}, R.~D., {Gosnell}, N.~M., \& {Leiner}, E.~M. 2023, \apj, 944, 145

\bibitem[{{Pancino} {et~al.}(2017){Pancino}, {Romano}, {Tang}, {Tautvai{\v{s}}ien{\.{e}}}, {Casey}, {Gruyters}, {Geisler}, {San Roman}, {Randich}, {Alfaro}, {Bragaglia}, {Flaccomio}, {Korn}, {Recio-Blanco}, {Smiljanic}, {Carraro}, {Bayo}, {Costado}, {Damiani}, {Jofr{\'e}}, {Lardo}, {de Laverny}, {Monaco}, {Morbidelli}, {Sbordone}, {Sousa}, \& {Villanova}}]{Pancinoetal2017}
{Pancino}, E., {Romano}, D., {Tang}, B., {et~al.} 2017, \aap, 601, A112

\bibitem[{{Panthi} \& {Vaidya}(2024)}]{PanthiVaidya2024}
{Panthi}, A. \& {Vaidya}, K. 2024, \mnras, 527, 10335

\bibitem[{{Parsons} {et~al.}(2018){Parsons}, {G{\"a}nsicke}, {Marsh}, {Ashley}, {Breedt}, {Burleigh}, {Copperwheat}, {Dhillon}, {Green}, {Hermes}, {Irawati}, {Kerry}, {Littlefair}, {Rebassa-Mansergas}, {Sahman}, {Schreiber}, \& {Zorotovic}}]{parsonsetal2018}
{Parsons}, S.~G., {G{\"a}nsicke}, B.~T., {Marsh}, T.~R., {et~al.} 2018, \mnras, 481, 1083

\bibitem[{{Philip Monai} {et~al.}(2024){Philip Monai}, {Martin}, \& {Jeffery}}]{PhilipMonetal2024}
{Philip Monai}, A., {Martin}, P., \& {Jeffery}, C.~S. 2024, \mnras, 527, 5408

\bibitem[{{Piotto} {et~al.}(2002){Piotto}, {King}, {Djorgovski}, {Sosin}, {Zoccali}, {Saviane}, {De Angeli}, {Riello}, {Recio-Blanco}, {Rich}, {Meylan}, \& {Renzini}}]{Piottoetal2002}
{Piotto}, G., {King}, I.~R., {Djorgovski}, S.~G., {et~al.} 2002, \aap, 391, 945

\bibitem[{{Portegies Zwart} {et~al.}(1999){Portegies Zwart}, {Makino}, {McMillan}, \& {Hut}}]{PortegiesZwartetal1999}
{Portegies Zwart}, S.~F., {Makino}, J., {McMillan}, S.~L.~W., \& {Hut}, P. 1999, \aap, 348, 117

\bibitem[{{Postnov} \& {Yungelson}(2014)}]{PostnovYoung2014}
{Postnov}, K.~A. \& {Yungelson}, L.~R. 2014, Living Reviews in Relativity, 17, 3

\bibitem[{{Prantzos} {et~al.}(2007){Prantzos}, {Charbonnel}, \& {Iliadis}}]{Prantzosetal2007}
{Prantzos}, N., {Charbonnel}, C., \& {Iliadis}, C. 2007, \aap, 470, 179

\bibitem[{{Rozner} \& {Perets}(2022)}]{RoznerPerets2022}
{Rozner}, M. \& {Perets}, H.~B. 2022, \apj, 931, 149

\bibitem[{{Salgado} {et~al.}(2022){Salgado}, {Da Costa}, {Yong}, {Salinas}, {Norris}, {Mackey}, {Marino}, \& {Milone}}]{Salgadoetal2022}
{Salgado}, C., {Da Costa}, G.~S., {Yong}, D., {et~al.} 2022, \mnras, 515, 2511

\bibitem[{{Schiavon} {et~al.}(2017){Schiavon}, {Zamora}, {Carrera}, {Lucatello}, {Robin}, {Ness}, {Martell}, {Smith}, {Garc{\'\i}a-Hern{\'a}ndez}, {Manchado}, {Sch{\"o}nrich}, {Bastian}, {Chiappini}, {Shetrone}, {Mackereth}, {Williams}, {M{\'e}sz{\'a}ros}, {Allende Prieto}, {Anders}, {Bizyaev}, {Beers}, {Chojnowski}, {Cunha}, {Epstein}, {Frinchaboy}, {Garc{\'\i}a P{\'e}rez}, {Hearty}, {Holtzman}, {Johnson}, {Kinemuchi}, {Majewski}, {Muna}, {Nidever}, {Nguyen}, {O'Connell}, {Oravetz}, {Pan}, {Pinsonneault}, {Schneider}, {Schultheis}, {Simmons}, {Skrutskie}, {Sobeck}, {Wilson}, \& {Zasowski}}]{Schiavonetal2017}
{Schiavon}, R.~P., {Zamora}, O., {Carrera}, R., {et~al.} 2017, \mnras, 465, 501

\bibitem[{{Scott} {et~al.}(2023){Scott}, {Jeffery}, {Farren}, {Tap}, \& {Dorsch}}]{Scottetal2023}
{Scott}, L.~J.~A., {Jeffery}, C.~S., {Farren}, D., {Tap}, C., \& {Dorsch}, M. 2023, \mnras, 521, 3431

\bibitem[{{Shariat} {et~al.}(2025){Shariat}, {Naoz}, {El-Badry}, {Rodriguez}, {Hansen}, {Angelo}, \& {Stephan}}]{Shariatetal2025}
{Shariat}, C., {Naoz}, S., {El-Badry}, K., {et~al.} 2025, \apj, 978, 47

\bibitem[{{Sills} {et~al.}(2005){Sills}, {Adams}, \& {Davies}}]{Sillsetal2005}
{Sills}, A., {Adams}, T., \& {Davies}, M.~B. 2005, \mnras, 358, 716

\bibitem[{{Sills} \& {Glebbeek}(2010)}]{Sillsgleb2010}
{Sills}, A. \& {Glebbeek}, E. 2010, \mnras, 407, 277

\bibitem[{{Simpson} {et~al.}(2017){Simpson}, {De Silva}, {Martell}, {Navin}, \& {Zucker}}]{Simpsonetal2017}
{Simpson}, J.~D., {De Silva}, G., {Martell}, S.~L., {Navin}, C.~A., \& {Zucker}, D.~B. 2017, \mnras, 472, 2856

\bibitem[{{Tang} {et~al.}(2021){Tang}, {Wang}, {Huang}, {Li}, {Yu}, {Geisler}, {Dias}, {Fern{\'a}ndez-Trincado}, {Carballo-Bello}, \& {Cabrera-Lavers}}]{Tangetal2021}
{Tang}, B., {Wang}, Y., {Huang}, R., {et~al.} 2021, \apj, 908, 220

\bibitem[{{Tokovinin}(2023)}]{Tokovinin2023}
{Tokovinin}, A. 2023, \aj, 165, 220

\bibitem[{{Tokovinin} {et~al.}(2006){Tokovinin}, {Thomas}, {Sterzik}, \& {Udry}}]{Tokovininetal2006}
{Tokovinin}, A., {Thomas}, S., {Sterzik}, M., \& {Udry}, S. 2006, \aap, 450, 681

\bibitem[{{Tylenda} {et~al.}(2024){Tylenda}, {Kami{\'n}ski}, \& {Smolec}}]{tylendaetal2024}
{Tylenda}, R., {Kami{\'n}ski}, T., \& {Smolec}, R. 2024, \aap, 685, A49

\bibitem[{{Wang} {et~al.}(2020){Wang}, {Kroupa}, {Takahashi}, \& {Jerabkova}}]{Wangetal2020}
{Wang}, L., {Kroupa}, P., {Takahashi}, K., \& {Jerabkova}, T. 2020, \mnras, 491, 440

\bibitem[{{Zhang} {et~al.}(2017){Zhang}, {Hall}, {Jeffery}, \& {Bi}}]{Zhangetal2017}
{Zhang}, X., {Hall}, P.~D., {Jeffery}, C.~S., \& {Bi}, S. 2017, \apj, 835, 242

\bibitem[{{Zhang} {et~al.}(2023){Zhang}, {Jeffery}, {Su}, \& {Bi}}]{Zhangetal2023}
{Zhang}, X., {Jeffery}, C.~S., {Su}, J., \& {Bi}, S. 2023, \apj, 959, 24

\end{thebibliography}

\end{document}